\documentstyle[aps,twocolumn]{revtex}

\begin{document}
\title{\begin{flushright}
\footnotesize{CECS-PHY-99/18}\\
\footnotesize{ULB-TH-99/26}
\end{flushright} Conserved Charges for Even Dimensional Asymptotically AdS
Gravity Theories}
\author{$^{2,4}$Rodrigo Aros\thanks{${\tt rod@cecs.cl}$}, $^{2,4}$Mauricio Contreras%
\thanks{${\tt contrera@cecs.cl}$}, $^{1,2}$Rodrigo Olea\thanks{${\tt %
rolea@cecs.cl}$}, $^{2,5}$Ricardo Troncoso\thanks{${\tt ratron@cecs.cl}$}, $%
^{2,3}$Jorge Zanelli\thanks{${\tt jz@cecs.cl}$}}
\address{$^{1}$ Departamento de F\'{\i }sica, FCFM, U. de Chile, Casilla 487-3\\
Santiago, Chile\\
$^{2}$ Centro de Estudios Cient\'{\i }ficos, Casilla 1469, Valdivia, Chile.\\
$^{3}$ Universidad de Santiago de Chile, Casilla 307, Santiago 2, Chile. \\
$^{4}$Universidad Nacional Andres Bello, Sazie 2320, Santiago, Chile.\\
$^{5}$Physique Th\'{e}orique et Math\'{e}matique, Universit\'{e} Libre de\\
Bruxelles, Campus Plaine, C.P.231, B-1050, Bruxelles, Belgium.}
\maketitle

\begin{abstract}
Mass and other conserved Noether charges are discussed for solutions of
gravity theories with locally Anti-de Sitter asymptotics in $2n$\
dimensions. The action is supplemented with a boundary term whose purpose is
to guarantee that it reaches an extremum on the classical solutions,
provided the spacetime is locally AdS at the boundary.\ It is also shown
that if spacetime is locally AdS at spatial infinity, the conserved charges
are finite and properly normalized without requiring subtraction of a
reference background. In this approach, Noether charges associated to
Lorentz and diffeomorphism invariance vanish identically for constant
curvature spacetimes. The case of zero cosmological constant is obtained as
a limit of AdS, where{\bf \ }$\Lambda $\ plays the role of a regulator.
\end{abstract}

\section{Introduction}

Noether's theorem is the standard tool in Theoretical Physics to construct
conserved charges associated with invariances of the action. Nevertheless,
General Relativity, described by Einstein-Hilbert action, does not lend
itself naturally to the application of Noether's theorem.\ The conserved
charge associated to the invariance of the action under diffeomorphisms is
given by Komar's formula \cite{Komar} 
\begin{equation}
K(\xi )=-\kappa \int_{\partial \Sigma }\nabla ^{\mu }\xi ^{\nu }d\Sigma
_{\mu \nu },  \label{komar}
\end{equation}
where $\kappa =(16\pi G)^{-1}$, $\xi =\xi ^{\mu }\partial _{\mu }$ is a
vector field that defines the diffeomorphism, $\nabla _{\mu }$ represents
the covariant derivative in terms of Christoffel symbol, $\partial \Sigma $
is the boundary of the spatial section, and $d\Sigma _{\mu \nu }=\frac{1}{2}%
\epsilon _{\mu \nu \alpha \beta }dx^{\alpha }\wedge dx^{\beta }$ is the
surface element (dual of the area two-form). Then, when $\xi $ is a timelike
or rotational Killing vector, $K(\xi )$ provides a definition of mass or
angular momentum, respectively. However, there is a first drawback in this
result, that is, in the case of the (3+1)-dimensional Kerr black hole,
equation (\ref{komar}) gives the following answer 
\begin{equation}
\begin{array}{ll}
K(\frac{\partial }{\partial t})=\frac{M}{2}; & K(\frac{\partial }{\partial
\phi })=J.
\end{array}
\label{MJ}
\end{equation}

These results show that there is no common normalization factor which could
give the correct values for mass and angular momentum.

Moreover, there is a second drawback with Komar's formula: in the presence
of negative cosmological constant, spacetime is no longer asymptotically
flat and the formula yields a divergent value. For example, for
Schwarzschild-AdS metric, one obtains 
\begin{equation}
\begin{array}{l}
K(\frac{\partial }{\partial t})=\frac{M}{2}+\lim\limits_{r\rightarrow \infty
}\frac{r^{3}}{2l^{2}}.
\end{array}
\label{komar-divergent}
\end{equation}

The standard approach to deal with this divergence is to subtract the value
of $K(\xi )$ on the AdS background from (\ref{komar-divergent}) (see, e.g. 
\cite{Julia-Silva}). In spite of giving a finite result, this does not
correct the normalization factor of $M$ and the first problem mentioned
above remains.

The usual procedure to evaluate the conserved charges is the ADM formalism 
\cite{ADM}, which yields the correct formulas for the energy-momentum and
angular momentum for asymptotically flat spacetimes. Nevertheless, this
approach and its further extension developed by Regge and Teitelboim \cite
{Regge-Teitelboim} provides a formula for the variation of the charges
--e.g., $\delta M$--, and in order to evaluate the charges --e.g., $M$--, it
is necessary to{\bf \ }fix the reference background geometry. The
Hamiltonian method can also be extended to provide the correct mass and
angular momentum for asymptotically AdS spacetimes representing solutions of
Einstein-Hilbert action with negative cosmological constant in $d=4$\ \cite
{Henneaux-Teitelboim}, as well as for $d\neq 4$\ \cite{Henneaux}.

In many instances this scheme is sufficiently satisfactory, but there are
some cases of physical interest in which the asymptotic behavior can be
difficult to assess, as in the case of asymptotically{\em \ locally }anti-de
Sitter ({\bf ALAdS}) spaces.

A formalism to define `conserved' charges in asymptotically AdS spaces was
proposed by Ashtekar and Magnon \cite{Ashtekar-Magnon}, who used conformal
techniques to construct the conserved quantities. This construction makes no
reference to an action, and yet reproduces the charges obtained by
Hamiltonian methods \cite{Henneaux-Teitelboim}.

Another scheme has been recently proposed by Balasubramanian and Kraus \cite
{BK} who use the Einstein-Hilbert action with Dirichlet boundary conditions
for the metric, supplemented by counterterms in order to ensure the
finiteness of the stress tensor derived by the quasilocal energy definition 
\cite{Brown-York}. By adding a finite series of local invariants of the
boundary geometry, the counterterm action regularizes that definition of
energy. This idea was subsequently extended to higher dimensions in \cite
{EJM}. A different, non-polynomial, expression has been given in \cite{Mann}%
, which reduces to the previous one in the infinite cosmological constant
limit.

In \cite{ACOTZ3+1}, an alternative construction is proposed which yields the
conserved charges in 3+1 dimensional General Relativity with negative
cosmological constant and does not need to specify the background provided
it is ALAdS.

\section{Noether Charges in 3+1 ALAdS Gravity}

The approach presented in \cite{ACOTZ3+1} leads to a properly defined,
convergent expression for the Noether charges in 3+1 dimensions, provided
the ALAdS boundary condition is imposed on the manifold. It is important to
note that the local AdS behavior at the boundary is not equivalent to the
usual Dirichlet condition over the metric, in order to have a well defined
variational principle.

The situation in 3+1 dimensions is reviewed in order to set the basic facts
in the construction. The starting observation is that the ALAdS condition
requires adding a boundary term to the Einstein-Hilbert action equal to the
Euler density (Gauss-Bonnet term) with a fixed weight factor, in order to
cancel the boundary term coming from the variation of the Lagrangian. As a
consequence, the action, including the boundary term, is

\begin{equation}
I=\frac{\kappa l^{2}}{4}\int\limits_{{\cal M}}\epsilon _{abcd}\bar{R}^{ab}%
\bar{R}^{cd}{},  \label{BI3+1}
\end{equation}
where $\bar{R}^{ab}:=R^{ab}+l^{-2}e^{a}e^{b}$. The\ Noether charge computed
with the action (\ref{BI3+1}) has the right normalization factor and is
finite for 3+1 dimensional ALAdS spaces. This charge, associated with the
invariance under a diffeomorphism of (\ref{BI3+1}), is 
\begin{equation}
Q(\xi )=\frac{\kappa l^{2}}{2}\int\limits_{\partial \Sigma }\epsilon
_{abcd}I_{\xi }\omega ^{ab}\bar{R}^{cd},  \label{q3+1}
\end{equation}
where $\xi ^{\mu }=x^{\prime \mu }-x^{\mu }$\ is the arbitrary vector field%
\footnote{%
The action of the contraction operator $I_{\xi }$ over a $p$-form $\alpha
_{p}=\frac{1}{p!}\alpha _{\mu ^{1}\cdot \cdot \cdot \mu ^{p}}dx^{\mu
^{1}}\!\cdot \cdot \cdot dx^{\mu ^{p}}$ is given by
\par
$I_{\xi }\alpha _{p}:=\frac{1}{(p-1)!}\xi ^{\nu }\alpha _{\nu \mu ^{1}\cdot
\cdot \cdot \mu ^{p-1}}dx^{\mu ^{1}}\!\cdot \cdot \cdot dx^{\mu ^{p-1}}$.
\par
In terms of this operator, the Lie derivative reads ${\cal L}_{\xi }=dI_{\xi
}+I_{\xi }d$ .} that generates the diffeomorphism.

Although Komar's formula and (\ref{q3+1}) are obtained as the conserved
Noether charge associated with\ the same invariance, they disagree because
the starting Lagrangians differ by a closed form and are deduced using
second and first order formalism, respectively. In order to clarify this
point, it is useful to split the charge (\ref{q3+1}) in such a way that the
relation with the usual tensor formalism becomes explicit. $Q(\xi )$\ can be
written as

\begin{equation}
Q(\xi )=K(\xi )+X(\xi )+\frac{\kappa l^{2}}{2}\int\limits_{\partial \Sigma
}\epsilon _{abcd}I_{\xi }\omega ^{ab}R^{cd},  \label{Qdecom}
\end{equation}
where $K(\xi )$\ is given by (\ref{komar}), $X(\xi )$\ is a contribution due
to the local Lorentz invariance\footnote{%
Here, the identity ${\cal L}_{\xi }e^{a}=D\xi ^{a}-I_{\xi }\omega
_{\,\,\,b}^{a}e^{b}$ , which holds in Riemannian (torsion-free) manifolds
has been used to obtain (\ref{Qdecom}).}, 
\begin{equation}
X(\xi )=-\frac{\kappa }{2}\int\limits_{\partial \Sigma }\epsilon _{abcd}\Phi
^{ab}e^{c}e^{d},  \label{Lorentzrotation}
\end{equation}
with $\Phi ^{ab}=e^{a\mu }{\cal L}_{\xi }e_{\mu }^{b}$\ and the last term
arises from the surface term in the action {\bf (}which was set as $\frac{%
\kappa l^{2}}{4}$ times Euler density). When $\xi $\ is a Killing vector, $%
\Phi ^{ab}$\ can be shown to be antisymmetric and be identified as a local
Lorentz transformation. In the second order formalism (\ref{Lorentzrotation}%
) is absent since there is no local Lorentz invariance\footnote{%
Although it is always possible to choose a frame $(e^{a})$\ such that $\Phi
^{ab}=0$\ in an open neighborhood, there could be interesting cases where a
global obstruction makes $X(\xi )$\ non-trivial.}.

The last term in (\ref{Qdecom}) plays a double role: it cancels the
divergences which appear in the explicit evaluation of the solutions and
contributes to the right normalization factor as well. In this sense, this
term\ regularizes the Noether charge for ALAdS spaces. This can be checked
explicitly in the following example: Consider the Schwarzschild-AdS solution
and $\xi =\partial _{t}$. In the standard frame choice $\Phi ^{ab}$ is zero
and hence $X(\xi )$\ vanishes. Evaluating (\ref{Qdecom}) yields 
\begin{equation}
K(\xi )=\frac{M}{2}+\lim_{r\rightarrow \infty }\frac{r^{3}}{2l^{2}}
\label{K}
\end{equation}
\begin{equation}
\frac{\kappa l^{2}}{2}\int\limits_{\partial \Sigma }\epsilon _{abcd}I_{\xi
}\omega ^{ab}R^{cd}=\frac{M}{2}-\lim_{r\rightarrow \infty }\frac{r^{3}}{%
2l^{2}},  \label{M/2}
\end{equation}
and hence, $Q(\xi )=M$.

It is apparent from relations (\ref{K}, \ref{M/2}), that\ the result (\ref
{Qdecom}) remains unchanged if the limit $l\rightarrow \infty $\ is taken at
the end. This permits applying the formula equally well for all values of
the cosmological constant, including $\Lambda =0$. In this sense, $\Lambda $
can be regarded as a regulator for General Relativity in the absence of
cosmological constant.

In what follows, the extension of this approach to $2n$-dimensional gravity
theories is presented. Since in higher dimensions, the Einstein-Hilbert (EH)
action is not the only option (see Section III below), we will also consider
a particular extension of the so-called Lanczos-Lovelock actions, which has
been dubbed the Born-Infeld (BI) action \cite{JJG}. This is an example that
the formalism can be applied to other theories of gravity that include
higher powers of curvature $R^{ab}$.

\section{Einstein-Hilbert Action}

\subsection{Action Principle}

In this section a well defined first order action principle for EH
Lagrangian in even dimensional ALAdS spacetimes is proposed. As in 3+1
dimensions{\bf , }the existence of an extremum for ALAdS spaces fixes\ the
boundary term that must be added to the action as proportional to the Euler
density. Applying of Noether's theorem to this action yields a regularized,
background-independent expression for the conserved charges.

The action to be considered is 
\begin{equation}
I=I_{EH}+B  \label{action}
\end{equation}
where $I_{EH}$ is the standard Einstein-Hilbert action with negative
cosmological constant in $d=2n$ dimensions, 
\begin{eqnarray}
I_{EH} &=&\frac{\kappa }{2(n-1)}\int\limits_{{\cal M}}\epsilon _{a_{1}\ldots
a_{d}}(R^{a_{1}a_{2}}e^{a_{3}}\ldots e^{a_{d}}  \nonumber \\
&&+\frac{d-2}{l^{2}d}e^{a_{1}}\ldots e^{a_{d}}),
\end{eqnarray}
and $B$ is a boundary term\footnote{%
Here, wedge product $\mbox{\tiny $\wedge$}$ between differential forms is
understood. The gravitational constant has been chosen as $\kappa =\frac{1}{%
2(d-2)!\Omega _{d-2}}$ with $\Omega _{d-2}$ the volume of $S^{d-2}$.}.

The on-shell variation of the action yields the boundary term 
\begin{equation}
\delta I=\int\limits_{\partial {\cal M}}\Theta ,  \label{deltaI}
\end{equation}
where 
\begin{equation}
\int\limits_{\partial {\cal M}}\Theta =\frac{\kappa }{(d-2)}%
\int\limits_{\partial {\cal M}}\epsilon _{a_{1}...a_{d}}\delta \omega
^{a_{1}a_{2}}e^{a_{3}}...e^{a_{d}}+\delta B.  \label{Theta}
\end{equation}
Therefore, the action becomes stationary demanding $\Theta =0$. Assuming the
spacetime to be ALAdS ($\bar{R}^{ab}=R^{ab}+l^{-2}e^{a}e^{b}=0$), the
vanishing of (\ref{Theta}) term is{\bf \ }satisfied if

\begin{equation}
\delta B=n\alpha _{n}\int\limits_{\partial {\cal M}}\!\!\!\!\epsilon
_{a_{1}...a_{2n}}\delta \omega ^{a_{1}a_{2}}R^{a_{3}a_{4}}\!\cdot \!\cdot
\!\cdot \!R^{a_{2n-1}a_{2n}}.  \label{deltaB}
\end{equation}
where $\alpha _{n}$ is defined as 
\begin{equation}
\alpha _{n}=\kappa \frac{(-1)^{n}l^{2n-2}}{2n(n-1)}.  \label{alphan}
\end{equation}

The r.h.s. of (\ref{deltaB}) can be recognized as the variation of the
2n-dimensional Euler density\footnote{%
Here we have defined the $2n$-dimensional Euler density as ${\cal E}%
_{2n}=\epsilon _{a_{1}...a_{2n}}R^{a_{1}a_{2}}...R^{a_{2n-1}a_{2n}}$. Note
that the normalization adopted here differs from standard mathematical
convention as, for instance, in \cite{Spivak}.}

\[
\delta {\cal E}_{2n}=n\int\limits_{{\cal M}}d\left[ \epsilon
_{a_{1}...a_{2n}}\delta \omega
^{a_{1}a_{2}}R^{a_{3}a_{4}}...R^{a_{2n-1}a_{2n}}\right] . 
\]
Thus, the boundary term in (\ref{action}) reads 
\[
B=\alpha _{n}\int\limits_{{\cal M}}{\cal E}_{2n}, 
\]
and the final expression for the action supplemented by the boundary term is 
\begin{equation}
I=I_{EH}+\alpha _{n}\int\limits_{{\cal M}}{\cal E}_{2n}.  \label{EH+Euler}
\end{equation}
This particular form of action is our starting point for the construction of
the conserved charges. The topology of the manifold is assumed to be ${\cal M%
}=R\times \Sigma $.

The diffeomorphism invariance is guaranteed by construction because the
action (\ref{EH+Euler}) is written in terms of differential forms. Thus,
Noether's theorem provides a conserved current (\ref{JNoether}) associated
with this invariance [see Appendix], given by 
\begin{equation}
\ast J=-\Theta (\omega ^{ab},e^{a},\delta \omega ^{ab})-I_{\xi }L,
\label{DifferomorphismCurrent}
\end{equation}
where $\delta \omega ^{ab}=-{\cal L}_{\xi }\omega ^{ab}$, and the Lagrangian 
$L$ can be read from (\ref{EH+Euler}). Then, $\Theta $ can be identified
from (\ref{Theta}) as 
\begin{eqnarray}
\Theta &=&-n\alpha _{n}\epsilon _{a_{1}...a_{2n}}{\cal L}_{\xi }\omega
^{a_{1}a_{2}}\left[ R^{a_{3}a_{4}}...R^{a_{2n-1}a_{2n}}\right.  \nonumber \\
&&+\left. (-1)^{n}\frac{e^{a_{3}}...e^{a_{2n}}}{l^{2n-2}}\right] .
\label{ThetaCondition}
\end{eqnarray}
The useful identity ${\cal L}_{\xi }\omega ^{ab}=DI_{\xi }\omega
^{ab}+I_{\xi }R^{ab}$ allows writing the conserved current (\ref
{DifferomorphismCurrent}) as an exact form. Thus, the conserved charge can
be written as 
\begin{eqnarray}
Q(\xi ) &=&n\alpha _{n}\int\limits_{\partial \Sigma }\epsilon
_{a_{1}...a_{2n}}I_{\xi }\omega ^{a_{1}a_{2}}\left[
R^{a_{3}a_{4}}...R^{a_{2n-1}a_{2n}}\right.  \nonumber \\
&&+\left. (-1)^{n}\frac{e^{a_{3}}...e^{a_{2n}}}{l^{2n-2}}\right] .
\label{TheconservedCharge}
\end{eqnarray}

This expression can also be written as 
\begin{equation}
Q(\xi )=\int\limits_{\partial \Sigma }I_{\xi }\omega ^{ab}{\cal T}_{ab},
\label{Q/Phi}
\end{equation}
where, ${\cal T}_{ab}$ is the variation of the Lagrangian with respect
Lorentz curvature 
\begin{equation}
{\cal T}_{ab}=\frac{\delta L}{\delta R^{ab}}.  \label{Phi}
\end{equation}

The general form adopted by the charge (\ref{Q/Phi}), can in fact be used
for any suitable gravitational theory --possessing a unique cosmological
constant--, whose Lagrangian is a polynomial in the curvature $R^{ab}$ and
the vielbein $e^{a}$, and has the right boundary terms to ensure the action
to have an extremum for ALAdS configurations.

It is noteworthy that this formula has been derived without making any
assumptions about a background geometry. The ALAdS condition restricts only
the\ local asymptotic relation between the curvature and the vielbein, with
no mention of the global topology of the manifold.

If $\xi $\ is a Killing vector globally defined on the boundary $\partial
\Sigma $, the surface integral (\ref{TheconservedCharge}) is the mass when $%
\xi =\partial _{t}$. Similarly, for other asymptotic Killing vectors, (\ref
{TheconservedCharge}) gives finite values for the linear and angular
momentum for a broad class of geometries. These statements are explicitly
checked below for different ALAdS spacetimes with inequivalent topologies.

\subsection{Examples}

\begin{itemize}
\item  {\bf Schwarzschild-AdS Black Hole}
\end{itemize}

The simplest example to be considered corresponds to the $d$-dimensional
black hole solution for the EH action with cosmological constant, known as
the Schwarzschild-AdS geometry, 
\begin{equation}
ds^{2}=-\Delta (r)^{2}dt^{2}+\frac{dr^{2}}{\Delta (r)^{2}}+r^{2}d\Omega
_{d-2}^{2},  \label{SphericalMetric}
\end{equation}
where $\Delta (r)^{2}=1-\frac{2M}{r^{d-3}}+\frac{r^{2}}{l^{2}}$.

The only non-vanishing charge is associated with the time-like Killing
vector $\partial _{t}$. Evaluating (\ref{TheconservedCharge}) on this metric
yields 
\begin{equation}
Q(\frac{\partial }{\partial t})=M.
\end{equation}

\begin{itemize}
\item  {\bf Kerr-AdS Solution}
\end{itemize}

In $d=2n$ dimensions, the rotating black hole solution is labeled by the
mass and $n-1$ parameters which are related to the Casimir invariants of $%
SO(d-1)$. The one parameter Kerr-AdS spacetime, representing a solution with
mass and angular momentum along a single axis, is given by the following
choice of vielbein \cite{HHTR}

\begin{equation}
\begin{array}{ll}
e^{0}=\Delta _{r}(dt-a\sin ^{2}(\theta )d\phi ) & e^{1}=\frac{1}{\Delta _{r}}%
dr \\ 
&  \\ 
e^{3}=\sin (\theta )\Delta _{\theta }(adt-(r^{2}+a^{2})d\phi ) & e^{2}=\frac{%
1}{\Delta _{\theta }}d\theta  \\ 
&  \\ 
e^{i}=r\cos (\theta )\tilde{e}^{i} & 
\end{array}
\end{equation}
where $i=5\ldots d$, $\tilde{e}^{i}$ is the vielbein for $S^{d-4}$ and 
\begin{equation}
\begin{array}{l}
\Delta _{r}^{2}=\frac{(r^{2}+a^{2})(1+\frac{r^{2}}{l^{2}})-2mr^{5-d}}{\Xi
^{2}\rho ^{2}} \\ 
\\ 
\Delta _{\theta }^{2}=\frac{1-\frac{a^{2}\cos ^{2}(\theta )}{l^{2}}}{\Xi
^{2}\rho ^{2}} \\ 
\\ 
\rho ^{2}=r^{2}+a^{2}\cos ^{2}(\theta ) \\ 
\Xi =1-\frac{a^{2}}{l^{2}}.
\end{array}
\end{equation}

This geometry has two non-vanishing Noether charges, one associated with the
time-like Killing vector $\partial _{t}$ and the rotational Killing vector $%
\partial _{\phi }$, respectively. For each dimension, the conserved charges
depend on the parameters $m$, $a$. For instance, in $6$ dimensions the mass
and angular momentum are given by 
\begin{equation}
\begin{array}{lr}
Q(\frac{\partial }{\partial t})=\frac{m}{\Xi ^{2}} & Q(\frac{\partial }{%
\partial \phi })=\frac{ma}{2\Xi ^{2}}.
\end{array}
\end{equation}

\begin{itemize}
\item  {\bf (Un)wrapped Brane Solution}
\end{itemize}

Unlike the Schwarzschild-AdS solution, where the spherical symmetry implies
a manifest AdS asymptotic behavior --not only locally, but globally at the
boundary--, another kind of ALAdS geometry in $d$ dimensions corresponds to
a brane solution with flat transverse space, 
\begin{equation}
ds^{2}=-\Delta (r)^{2}dt^{2}+\frac{dr^{2}}{\Delta (r)^{2}}%
+r^{2}(dx_{1}^{2}+\ldots +dx_{d-2}^{2}),  \label{ToroidalMetric}
\end{equation}
where $\Delta (r)^{2}=-\frac{2m}{r^{d-3}}+\frac{r^{2}}{l^{2}}$. In this
geometry at least one of the $x^{i}$ coordinates must be compact, otherwise
the parameter $m$ can be absorbed by a coordinate transformation \cite
{HorowitzMyers}. Assuming that the volume of the transverse space ($x^{i}$)
is equal to $V$, the Noether charge associated with the Killing vector $%
\partial _{t}$ is given by 
\begin{equation}
Q(\frac{\partial }{\partial t})=m\frac{V}{\Omega _{d-2}},
\label{EHpbraneCharge}
\end{equation}
and the corresponding charges related{\bf \ }with spatial $ISO(d-2)$\
symmetries are zero.

It should be noted that (\ref{EHpbraneCharge}) depends on the topological
nature of the transverse spatial section. In the case the transverse space
is compact, $V$ is finite and so is the resulting Noether charge. When the
transverse space is non compact, the parameter $m$ can be interpreted as a
mass density.

This method provides the correct results even for the electrically charged
extensions of the previous solutions. It is straightforward to prove that
the formula works properly for the higher dimensional Reissner-Nordstr\"{o}m
black hole, for the $(3+1)$-dimensional Kerr-Newman solution, and for the
electrically charged generalization of the (un)wrapped brane (\ref
{ToroidalMetric}) studied in \cite{Cai-Zhang}.

\section{Born-Infeld Action}

In higher dimensions, besides the Einstein-Hilbert action one can consider
other gravitational theories that include higher powers of the curvature and
still yield second order field equations for the metric. Among them, there
are a few that lead to well behaved ALAdS solutions \cite{TZ}. In even
dimensions $(d=2n)$, the Born-Infeld ({\bf BI}) action belongs to this class
of theories. The BI action takes the explicit form 
\begin{equation}
I=\frac{\kappa l^{2}}{2n}\int\limits_{{\cal M}}\epsilon _{a_{1}\ldots a_{d}}%
\bar{R}^{a_{1}a_{2}}\ldots \bar{R}^{a_{d-1}a_{d}},  \label{BI_even_D}
\end{equation}
where $\bar{R}^{ab}=R^{ab}+l^{-2}e^{a}e^{b}$. This theory is stationary for
ALAdS solutions and no boundary terms are required in order to have a well
defined action principle\footnote{%
This expression can also be written as 
\[
L=2^{n-1}(n-1)!\kappa l^{2}\sqrt{\det (R^{ab}+l^{-2}e^{a}e^{b})}, 
\]
and for this reason it has been dubbed the Born-Infeld action \cite{JJG}. In
four dimensions, this action reduces to the usual Einstein-Hilbert with
cosmological constant plus the Euler density, with all coefficient fixed as
in previous section.}.

Following the Hamiltonian method, it would be extremely difficult to obtain
a mass formula as a surface integral for an arbitrary localized matter
distribution in this kind of theories. Such construction would require
inverting the symplectic matrix of this action. However, the rank of this
matrix depends on the fields and therefore no general form can be found for
an arbitrary field configuration. On the other hand, in the Lagrangian
formalism, the Noether current for diffeomorphisms is an exact form, which
allows writing down the conserved charge at once as the surface integral 
\begin{equation}
Q(\xi )=\int\limits_{\partial \Sigma }I_{\xi }\omega ^{ab}{\cal T}_{ab},
\label{BICharge}
\end{equation}
where 
\begin{equation}
{\cal T}_{ab}=\frac{\kappa l^{2}}{2}\epsilon _{aba_{3}\ldots a_{d}}\bar{R}%
^{a_{3}a_{4}}\ldots \bar{R}^{a_{d-1}a_{d}}.
\end{equation}
This is an appropriate definition of mass and other conserved charges, as is
shown in the following examples.

\begin{itemize}
\item  {\bf Static Spherically Symmetric Solution}
\end{itemize}

The spherically symmetric black hole solution of BI action was studied in 
\cite{BTZd}. In 3+1 dimensions this is the Schwarzschild-AdS geometry, but
differs from it in higher dimensions. The line element is given by 
\begin{equation}
ds^{2}=-\Delta (r)^{2}dt^{2}+\frac{dr^{2}}{\Delta (r)^{2}}+r^{2}d\Omega
_{d-2}^{2},
\end{equation}
with $\Delta (r)^{2}=1-\left( \frac{2M}{r}\right) ^{\frac{1}{n-1}}+\frac{%
r^{2}}{l^{2}}$.

The mass is obtained by direct computation of (\ref{BICharge}) for the
Killing vector $\xi =\partial _{t}$, 
\begin{equation}
Q(\frac{\partial }{\partial t})=M.
\end{equation}
The conserved charges associated with rotational isometries vanish.

\begin{itemize}
\item  {\bf (Un)wrapped Brane Solution}
\end{itemize}

A feature in common for the BI and EH actions is possessing a set of
solutions that are only ALAdS, but not globally AdS at the boundary. Among
many solutions, it is interesting to consider the analog of the (un) wrapped
brane (\ref{ToroidalMetric}), for which the line element reads 
\begin{equation}
ds^{2}=-\Delta (r)^{2}dt^{2}+\frac{dr^{2}}{\Delta (r)^{2}}%
+r^{2}(dx_{1}^{2}+\ldots +dx_{d-2}^{2}),
\end{equation}
with $\Delta (r)^{2}=-\left( \frac{2m}{r}\right) ^{\frac{1}{n-1}}+\frac{r^{2}%
}{l^{2}}$. This corresponds to a particular case of the class of geometries
studied by Cai and Soh in \cite{Cai-Soh}. The transverse space in this
gravitational configuration is (locally) flat, with a volume $V$. This
geometry has just one non-vanishing Noether charge, that is the density of
mass associated with the Killing vector $\partial _{t}$%
\begin{equation}
Q(\frac{\partial }{\partial t})=m\frac{V}{\Omega ^{d-2}}.
\label{BIpbraneCharge}
\end{equation}

This last result is in complete agreement with the one computed by the
Hamiltonian method, using a mini-superspace model applied to configurations
with transverse space not necessarily compact. However, this result differs
by a global factor compared to the same case as treated in \cite{Cai-Soh}%
{\it . }The origin of this mismatch lies in the fact that transverse space
is no longer spherically symmetric; therefore, the volume $V$ cannot cancel
the normalization factor, fixed beforehand to give the correct value of mass
for spherically symmetric black holes.

\section{Conserved Charge for Lorentz Transformations}

Apart from charges associated with diffeomorphisms and due to the invariance
of the EH and BI actions under local Lorentz transformations, the Noether
method can also be applied to obtain conserved quantities for these
symmetries [see Appendix]. Substituting $\delta \omega ^{ab}=-D\lambda ^{ab}$
in the general expression for the Noether current (\ref{JNoether})

\begin{equation}
\ast J=\delta \omega ^{ab}{\cal T}_{ab}\text{ },
\end{equation}
where ${\cal T}_{ab}$ is covariantly constant, yields the conserved charge
in terms of the parameter of the Lorentz transformation $\lambda ^{ab}$ as

\begin{equation}
Q(\lambda ^{ab})=\int\limits_{\partial \Sigma }\lambda ^{ab}{\cal T}_{ab}.
\label{LorentzVariationofDiiffChrge}
\end{equation}
Here

\begin{eqnarray}
{\cal T}_{ab} &=&n\alpha _{n}\epsilon _{aba_{3}...a_{2n}}\left[
R^{a_{3}a_{4}}...R^{a_{2n-1}a_{2n}}\right.  \nonumber \\
&\text{ }&\text{ }+\left. (-1)^{n}\frac{e^{a_{3}}...e^{a_{2n}}}{l^{2n-2}}%
\right] ,  \label{EHLorentzCharge}
\end{eqnarray}
for Einstein-Hilbert case, and

\begin{equation}
{\cal T}_{ab}=\frac{\kappa l^{2}}{2}\epsilon _{aba_{3}\ldots a_{d}}\bar{R}%
^{a_{3}a_{4}}\ldots \bar{R}^{a_{d-1}a_{d}},  \label{BILorentzCharge}
\end{equation}
for Born-Infeld action.

\begin{center}
{\bf Lorentz Covariance}
\end{center}

The formula (\ref{LorentzVariationofDiiffChrge}) is a scalar from the point
of view of Lorentz covariance. On the other hand, the charges (\ref{BICharge}%
) and (\ref{TheconservedCharge}) associated with diffeomorphism invariance
transform under local Lorentz rotations as

\begin{equation}
\delta _{\lambda }Q(\xi )=-\int\limits_{\partial \Sigma }{\cal L}_{\xi
}\lambda ^{ab}{\cal T}_{ab}.  \label{DeltaQ}
\end{equation}

This change in $Q(\xi )${\bf \ }vanishes under the usual assumption that the
local transformation with parameter $\lambda ^{ab}$ approaches a rigid
Lorentz transformation on $\partial \Sigma $, constant along $\xi $, that
is, ${\cal L}_{\xi }\lambda ^{ab}|_{\partial \Sigma }=0${\it .}

\section{Summary and Prospects}

The method presented here is the direct application of Noether's theorem to
a first order gravitational action $I[e,\omega ]$, provided the spacetime
satisfies ALAdS boundary conditions. The analysis leads directly to general
analytic expressions for the conserved charges, both for the
Einstein-Hilbert and Born-Infeld actions. The treatment is entirely
Lagrangian and yields values for the charges that match exactly those
obtained by Hamiltonian methods (e.g., ADM). In the Hamiltonian approach,
however, when the space is not asymptotically flat it is often necessary to
renormalize the asymptotic Killing vectors to define the conserved charges
(see, for example, \cite{BHTZ}).

The resulting charges are finite for localized distributions of matter
(black holes) and yield finite density formulas for extended objects (e.g.,
strings). There is no need to subtract the ``vacuum'' energy in order to
regularize the charges. It could be argued that the Euler density added as a
boundary term does this job for us, but what is indisputable is the fact
that one does not need to specify a reference background against which one
should compute the value of the charges. What could be even more surprising
is the fact that the formulas (\ref{TheconservedCharge}), for EH action, and
(\ref{BICharge}) for BI the case, give the correct answers even for
radically different asymptotic behaviors.

The general nature of the treatment, allows extending to $2n$ dimensions,
and also to the BI action, the following result valid for the EH action in
(3+1)-dimensions with negative cosmological constant: Noether charges
associated to Lorentz and diffeomorphism invariance vanish identically in
locally AdS spacetimes.

As can be directly checked from (\ref{TheconservedCharge}) and (\ref
{LorentzVariationofDiiffChrge}), the charges are identically zero if 
\[
\bar{R}^{ab}=R^{ab}+l^{-2}e^{a}e^{b}=0 
\]
in the bulk. This means that spaces which are locally AdS have vanishing
charges. In particular, any locally AdS geometry with a timelike Killing
vector should have zero mass \footnote{%
It should be stressed that this assertion is only valid in even dimensions,
for it is well known that at least in three dimensions different locally AdS
spaces can have different energies ($m=-1$ for AdS, $m\geq 0$ for black
holes \cite{BTZ}). This probably means that the analysis presented here
cannot be repeated {\em verbatim} in odd dimensions.}. This brings in an
interesting issue: there could be several topologically different spaces
with locally AdS geometry for which all their quantum numbers associated to
spatial transformations vanish. Each of these spaces could be reasonably
used as vacuum for a quantum field theory and one should also expect to find
interpolating instanton or soliton configurations. Also, any massive
solution such as the examples discussed above could be seen as an excitation
of the corresponding background in the same topological sector.

\begin{center}
{\bf Prospects}
\end{center}

Only two cases (EH and BI) have been considered here among all the possible
Lanczos-Lovelock theories of gravity \cite{Lanczos,Lovelock,Zumino}. The
suitable theories describing gravitation in higher dimensions must posses a
unique cosmological constant and therefore, a unique background in each
topological sector, so that vacuum configurations approach to local AdS
spacetimes with a fixed curvature radius at the boundary \cite{TZ}. Indeed,
there exists a subset of these theories which possesses well behaved black
hole solutions \cite{Scan}. The extension of this formalism to those
theories, in even dimensions, will be discussed elsewhere.

As mentioned above, the odd-dimensional manifolds cannot be\ treated with
the same method presented here. The cases of interest in $(2n+1)$
dimensions, analogous to those discussed in this note, would be EH and
Chern-Simons. Regularization of the charges in these cases remains an open
problem in the presented framework.

Another interesting problem to address is the classification of all $2n$%
-dimensional constant curvature spaces, as they can be thought of as
candidates for vacuum configurations for an AdS field theory. Certainly, one
possible class of such spaces could be AdS with identifications along global
Killing vectors (that do not introduce causal or conical singularities), but
it is not obvious that this exhausts all possibilities in high enough
dimensions.

\section{Appendix}

\subsection{Noether Theorem}

In order to fix the notation and conventions, here we briefly review
Noether's theorem.

Consider a $d$-form Lagrangian $L(\varphi ,d\varphi )$, where $\varphi $
denotes collectively a set of $p$-form fields. An arbitrary variation of the
action under a local change $\delta \varphi $ is given by the integral of 
\begin{equation}
\delta L=(E.O.M)\delta \varphi +d\Theta (\varphi ,\delta \varphi ),
\label{EOM}
\end{equation}
where E.O.M. stands for equations of motion and $\Theta $ is a corresponding
boundary term \cite{Ramond}. The total change in $\varphi $ ($\bar{\delta}%
\varphi =\varphi ^{\prime }(x^{\prime })-\varphi (x)$) can be decomposed as
a sum of a local variation and the change induced by a diffeomorphism, that
is, $\bar{\delta}\varphi =\delta \varphi +{\cal L}_{\xi }\varphi $, where $%
{\cal L}_{\xi }$ is the Lie derivative operator. In particular, a symmetry
transformation acts on the coordinates of the manifold as $\delta x^{\mu
}=\xi ^{\mu }(x)$, and on the fields as $\delta \varphi $, leading a change
in the Lagrangian given by $\delta L=d\Omega .$

Noether's theorem states that there exists a conserved current given by 
\begin{equation}
\ast J=\Omega -\Theta (\varphi ,\delta \varphi )-I_{\xi }L,  \label{JNoether}
\end{equation}
which satisfies $d*J=0$. This, in turn, implies the existence of the
conserved charge 
\[
Q=\int\limits_{\Sigma }{}*J, 
\]
where $\Sigma $ is the spatial section of the manifold, when a manifold is
assumed to be of topology $R\times \Sigma $ .

\section{Acknowledgments}

The authors are grateful to V. Balasubramanian, J. Cris\'{o}stomo, C. Mart\'{%
\i }nez, F. M\'{e}ndez, M. Plyushchay and C. Teitelboim for many
enlightening discussions and helpful comments. We are specially thankful to
M. Henneaux for many helpful comments. This work was supported in part
through grants 1990189, 1970151, 1980788, 2990018, 3960007, 3960009 and
3990009 from FONDECYT, and Grants No. DI 51-A/99 (UNAB), and 27-953/ZI-DICYT
(USACH), and by the ``Actions de Recherche Concert{\'{e}}es'' of the
``Direction de la Recherche Scientifique - Communaut{\'{e}} Fran{\c{c}}aise
de Belgique'', by IISN - Belgium (convention 4.4505.86). The institutional
support of Fuerza A\'{e}rea de Chile, I.\ Municipalidad de Las Condes, and a
group of Chilean companies (AFP Provida, Business Design Associates, CGE,
CODELCO, COPEC, Empresas CMPC, GENER\ S.A., Minera Collahuasi, Minera
Escondida, NOVAGAS and XEROX-Chile) is also recognized. Also R.A., R.O. and
J.Z. wish to thank to the organizers of the ICTP Conference on Black Hole
Physics for hospitality in Trieste this last summer. CECS is a Millennium
Science Institute.

\end{document}